\documentclass[iop]{emulateapj}
\usepackage{natbib,graphics,graphicx}

\newcommand{\snr}{Kes 17}
\begin{document}
\title{Supernova Remnant Kes 17: Efficient Cosmic Ray Accelerator
  inside a Molecular Cloud} 
\author{Joseph D. Gelfand}
\affil{NYU Abu Dhabi}
\affil{PO Box 903, New York, NY 10276}
\email{jg168@cosmo.nyu.edu}
\altaffilmark{1}
\altaffiltext{1}{Affiliate Member, Center for Cosmology and Particle
  Physics, New York University}
\and
\author{Daniel Castro}
\affil{Kavli Institute for Astrophysics \& Space Research,
  Massachusetts Institute of Technology}
\and
\author{Patrick O. Slane}
\affil{Harvard-Smithsonian Center for Astrophysics}
\affil{60 Garden Street, Cambridge, MA 02138}
\and
\author{Tea Temim}
\affil{Observational Cosmology Lab, Code 665}
\affil{NASA Goddard Space Flight Center}
\affil{Greenbelt, MD 20771, USA}
\altaffilmark{2}
\altaffiltext{2}{Oak Ridge Associated Universities, Oak Ridge, TN
  37831, USA} 
\and
\author{John P. Hughes}
\affil{Department of Physics and Astronomy}
\affil{Rutgers University}
\affil{136 Frelinghuysen Rd, Piscataway, NJ 08854, USA}
\and
\author{Cara Rakowski}
\affil{United States Patent and Trademark Office}
\affil{600 Dulany Street, Alexandria, VA}
\email{cara.rakowski@gmail.com}

\begin{abstract}
  Supernova remnant Kes 17 (SNR G304.6+0.1) is one of a few but
  growing number of remnants detected across the electromagnetic
  spectrum.  In this paper, we analyze recent radio, X-ray, and
  $\gamma$-ray observations of this object, determining that efficient
  cosmic ray acceleration is required to explain its broadband
  non-thermal spectrum.  These observations also suggest that Kes 17
  is expanding inside a molecular cloud, though our determination of
  its age depends on whether thermal conduction or clump evaporation
  is primarily responsible for its center-filled thermal X-ray
  morphology.  Evidence for efficient cosmic ray acceleration in Kes
  17 supports recent theoretical work that the strong magnetic field,
  turbulence, and clumpy nature of molecular clouds enhances cosmic
  ray production in supernova remnants.  While additional observations
  are needed to confirm this interpretation, further study of Kes 17
  is important for understanding how cosmic rays are accelerated in
  supernova remnants.
\end{abstract}

\section{Introduction}
\label{intro}

Supernova remnants (SNRs) are believed to be important in both forming
and regulating the multi-phase interstellar medium (ISM) found inside
star-forming galaxies (e.g., \citealt{mckee77}): distributing metals
produced in the progenitor explosion throughout the host galaxy,
producing dust (e.g., \citealt{saltpeter77}), and accelerating cosmic
rays up to energies $E\sim10^{15.5}~{\rm eV}$ (e.g.,
\citealt{arnett73}).  However, direct observational evidence
supporting this last assertion is rare.  While several Milky Way SNRs
are identified as cosmic ray producers, only one (Tycho's SNR) shows
evidence for accelerating protons up to the ``knee'' in the cosmic ray
spectrum believed to delineate Galactic from extragalactic cosmic rays
\citep{eriksen11}.

Determining if SNRs are responsible for the observed cosmic ray
population requires studying individual remnants to determine both if
and how they accelerate cosmic rays.  SNRs are extremely complicated
objects comprised of hot ISM material shocked by the expanding
supernova blast wave (the ``forward shock''), supernova ejecta heated
by the shock wave driven into the SNR by the shocked ISM (the
``reverse shock''), cold unshocked ejecta, and relativistic electrons
and ions accelerated at the forward and/or reverse shock.
Additionally, the dynamical evolution of the SNR depends strongly on
its surroundings (e.g., \citealt{lozinskaya92}).  Determining the
properties of relativistic particles accelerated inside an SNR
requires first measuring the physical properties of these different
components.

This requires analyzing an SNR's emission across the entire
electromagnetic spectrum.  An SNR's radio emission traces GeV
electrons accelerated in the remnant.  Its infrared (IR) emission is
produced by dust and atomic and molecular gas inside and outside the
SNR heated by shocks and higher energy emission.  A remnant's thermal
X-ray emission traces both ISM material shocked by the forward shock
and ejecta shocked by the reverse shock.  Lastly, its $\gamma$-ray
emission traces relativistic electrons, and possibly hadrons (cosmic
rays), accelerated in the SNR (e.g., \citealt{ackermann13}).  Thanks
to new observing capabilities at both IR wavelengths (e.g., {\it
  Spitzer}, {\it AKARI}, and {\it Herschel}) and $\gamma$-ray energies
(e.g., {\it Fermi}, H.E.S.S, V.E.R.I.T.A.S, and MAGIC), the number of
SNRs detected in all four wavebands is rapidly increasing.  One such
remnant is Kes 17 (SNR G304.6+0.1).  In this paper, we analyze recent
radio, IR, X-ray, and $\gamma$-ray observations of this remnant
(\S\ref{observations}) and use these results to determine the physical
properties of this SNR and its surroundings (\S\ref{interpretation}).
Finally, we summarize our results and discuss their implications on
the interaction between SNRs and their environments
(\S\ref{conclusions}).

\section{Observations and Data Analysis}
\label{observations}
In this Section, we analyze recent radio (\S\ref{radio}), IR
(\S\ref{ir}), X-ray (\S\ref{xray}), and $\gamma$-ray (\S\ref{gray})
observations of this source.

\begin{figure}[tb]
\begin{center}
\includegraphics[width=0.95\columnwidth,angle=0]{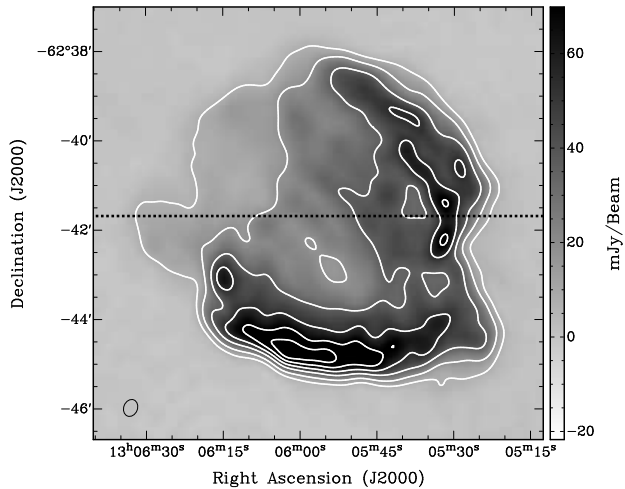}
\end{center}
\vspace*{-0.5cm}
\caption{1.4 GHz image of Kes 17.  The total intensity image was made
  from the combined data described in the \S\ref{radio} using uniform
  weighting, multi-frequency synthesis, and maximum entropy
  deconvolution.  This image has an rms noise of 0.72~mJy~beam$^{-1}$
  and a resolution of $23\farcs3\times18\farcs4$ (size and orientation
  of beam represented by the ellipse in the lower left-hand corner.)
  The white contours indicate surface brightness levels of 10, 25, 50,
  75, and 100 $\sigma$.  The dashed line indicates the region used to
  make the brightness profile shown in Figure \ref{fig:20cmprof}.}
\label{fig:20cmimg}
\end{figure}

\subsection{Radio}
\label{radio}
Kes 17 was first detected at 408 MHz and 5 GHz by \citet{shaver70}
who, based on the non-thermal spectrum implied by its flux at these
two frequencies, classified this source as an SNR.  This
identification was supported by the detection of polarized 5~GHz
emission \citep{milne75} and the irregular shell-like morphology
revealed by analysis of data taken during the Molonglo Observatory
Synthesis Telescope (MOST) Galactic plane survey \citep{whiteoak96}.
Analysis of the \ion{H}{1} spectrum using the standard \ion{H}{1}
absorption method towards Kes 17 suggests a distance $d>9.7~{\rm kpc}$
\citep{caswell75}.  Last, but not least, the detection of OH
(1720~MHz) maser emission \citep{frail96} requires the presence of
shocked molecular material (\citealt{elitzur76}, see
\citealt{wardle12} for a recent review).

\begin{table}[tb]
\begin{center}
\begin{tabular}{ccc}
\hline
\hline
{\sc Frequency} [GHz] & {\sc Flux Density} [Jy] & {\sc Reference} \\
\hline
0.408 & 29.8 & \citet{shaver70} \\
0.843 & 18 & \citet{whiteoak96} \\
1.4 & $10.9\pm0.14$ & This work \\
5.0 & 6.7 & \citet{shaver70} \\
\hline
\hline
\end{tabular}
\end{center}
\vspace*{-0.5cm}
\caption{Flux Densities of Kes 17 at several radio frequencies.
\label{tab:radiofluxes}} 
\end{table}

The Australia Telescope Compact Array, while in its 1.5A
configuration, observed this SNR on 2004 March 14 at both 1.4 \&
2.4~GHz.  This observation used the correlator setting with the
maximum bandwidth available (128~MHz bands over 13 channels), with one
band centered at 1384~MHz and the other at 2368 MHz.  This observation
recorded all four linear polarization modes (XX, YY, XY, and YX).  We
used the {\tt MIRIAD} software package \citep{sault95} to calibrate
the flux density using an observation of PKS B1934-638, calibrate the
phase using data from regular observations of PKS 1329$-$665, and
image the Kes 17 data.  To improve our sensitivity to diffuse 1.4 GHz
emission in the field, we combined the 1.4 GHz visibilities of Kes 17
with continuum data from the Southern Galactic Plane Survey
\citep{sgps}.

\begin{figure}[tb]
\begin{center}
\includegraphics[width=0.85\columnwidth,angle=270]{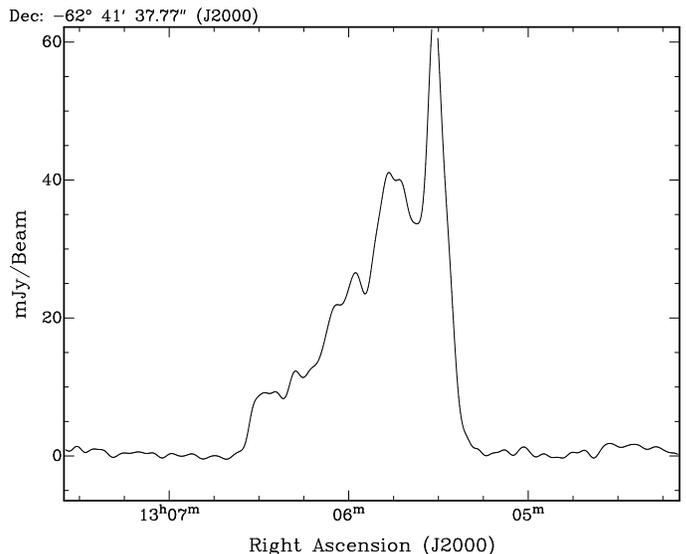}
\end{center}
\vspace*{-0.5cm}
\caption{1.4 GHz intensity profile of Kes 17 along a line of
  constant declination $\delta=-62^\circ 41^\prime 37\farcs77$
  (J2000).  At this declination, Kes 17 has the largest angular
  extent.}
\label{fig:20cmprof}
\end{figure}

As shown in Figure \ref{fig:20cmimg}, at 1.4 GHz this SNR has a
partial shell morphology with a diameter of $\sim7\farcm5$ dominated
by two rims in the S and NW regions connected by a``notch''-like
feature in the SW.  While the NE region has a surface brightness
$\sim6\times$ lower than the S and NW rims, there is a sharp decrease
in flux density that defines the edge of this remnant (Figure
\ref{fig:20cmprof}).  Diffuse radio emission is also detected interior
to the shell (Figures \ref{fig:20cmimg} \& \ref{fig:20cmprof}).  The
total (Stokes I) 1.4 GHz flux density of Kes 17 is $10.9\pm0.14$~Jy
and no polarized emission was detected.  The lack of data with short
{\it u--v} spacing at 2.4 GHz precluded making a similar quality
image and flux density measurement at this higher frequency.

\subsection{Infrared}
\label{ir}
The first IR detection of Kes 17 was made with the {\it Infrared
  Astronomical Satellite} ({\it IRAS}), which revealed shell-like
emission from this SNR \citep{arendt89}. More recent \textit{Spitzer}
Infrared Array Camera (IRAC) observations from the GLIMPSE survey
uncovered very bright emission in the 3.6-8.0 $\micron$ bands
\citep{lee05, reach06} from a more diffuse SNR shell.  The filamentary
structure along the NW rim is particularly bright at 4.5 $\micron$,
suggesting the emission originates from shocked H$_2$. Based on the
colors and morphological similarities of the IRAC images,
\citet{reach06} concluded this emission is produced by molecular
shocks.

Analysis of 5--38 $\micron$ spectroscopic follow-up observations by the
Infrared Spectrograph (IRS) aboard \textit{Spitzer} revealed bright
pure rotational lines of H$_2$, most likely indicating an interaction
between the SNR and dense molecular material \citep{hewitt09}.  Shock
models suggest the excitation of the observed H$_2$ lines requires two
shock components; a slower 10 ${\rm km\:s}^{-1}$ C-shock through
denser clumps with $n_0=10^6\: {\rm cm}^{-3}$, and a faster 40 ${\rm
  km\:s}^{-1}$ C-shock passing through a lower density medium with
$n_0=10^4\: {\rm cm}^{-3}$ \citep{hewitt09}.  Analysis of the spectra
also reveal atomic fine-structure lines of \ion{Fe}{2}, \ion{Ne}{2},
\ion{Ne}{3}, \ion{S}{3}, \ion{S}{1}, and \ion{S}{2}, whose relative
emission line fluxes lead to densities in the 100 -- 1000 ${\rm
  cm}^{-3}$ range and shock velocities of 150 -- 200 ${\rm
  km\:s}^{-1}$ \citep{hewitt09}.

Most recently, the broadband mid to far-IR emission from Kes 17 was
detected by the Multiband Imaging Photometer for \textit{Spitzer}
(MIPS) at 24 $\micron$ \citep{lee11,pinheiro11} and the \textit{AKARI}
satellite at 15, 24, 65, 90, 140, and 160 $\micron$ \citep{lee11}.
Emission at these wavelengths is concentrated in the W and S shells,
partially overlapping with the W radio rim. The broadband IR spectral
energy distribution is well fit by two modified blackbodies with a
mixture of carbonaceous and silicate dust grain compositions. The best
fit temperatures are $79\pm6\:\rm K$ and $27\pm3\:\rm K$ with dust
masses of $6.2\times10^{-4}\: \rm M_{\odot}$ for the hot component and
$6.7\: \rm M_{\odot}$ for the cold component for a distance of 8 kpc
\citep{lee11}.  While this distance is modestly inconsistent with that
estimated from the \ion{H}{1} absorption spectrum of this SNR
(\S\ref{radio}), this discrepancy does not significantly change these
masses.

\subsection{X-ray}
\label{xray}

X-ray emission from Kes 17 was first detected in an unpublished
$\sim11$~ks {\it ASCA} observation ({\sc ObsID} 57013000) on 1999
February 12.  {\it XMM-Newton} then observed this SNR on 2005 August
12 ({\sc ObsID} 0303100201) for $\sim20$~ks after the removal of
background flares \citep{combi10}.  Analysis of this observation
revealed the X-ray emission was diffuse and extended, brightest inside
the radio shell, and suggested the presence of both non-thermal and
thermal X-ray emission components and spatial variations in its X-ray
spectrum \citep{combi10}.

More recently, Kes 17 was observed on 2010 September 3 by the {\it
  Suzaku} observatory for $\sim100$~ks ({\sc ObsID} 505074010).  We
first reprocessed the {\it Suzaku} data using the new {\sc aepipeline}
task in the {\tt ftools v6.11} software package\footnote{Available at
  http://heasarc.gsfc.nasa.gov/ftools/} \citep{blackburn95}, then used
a circular region $4\farcm7$ in radius centered on Kes 17 to extract
its spectrum, determining the background spectrum using data from an
annulus between $4\farcm7$ and $7\farcm4$ in radius also centered on
Kes 17.  For consistency, we used the same source and background
region for all three detectors.  The spectra were created using {\tt
  xselect}, as were the RMF and ARF of the source spectrum.  We then
fit the observed, background subtracted 0.5--10~keV spectrum to
different emission models using the {\tt Sherpa} \citep{freeman01}
software package, checking our results with {\tt XSpec}
\citep{arnaud96} since these packages use different algorithms to
determine the best fit and errors on the model parameters.

\begin{figure}[tb]
\begin{center}
\includegraphics[width=0.95\columnwidth]{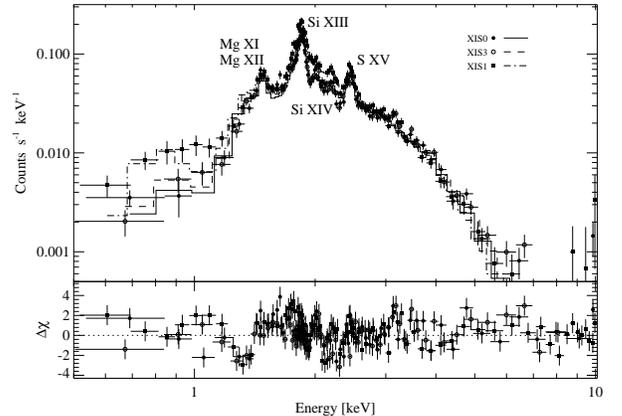}
\end{center}
\vspace*{-0.5cm}
\caption{The {\it Suzaku} X-ray spectrum of Kes 17, overlaid with the
  prediction of the {\tt phabs $\times$ vnei} model given in Table
  \ref{tab:xrayspec}.}
\label{fig:model1}
\end{figure}

\begin{table*}[tb]
\begin{center}
\small
\begin{tabular}{ccccccc}
\hline
\hline
{\sc Parameters} & {\tt vnei} & {\tt vray} & {\tt vgnei}  & {\tt
  vsedov} & {\tt vpshock} & {\tt vnpshock} \\
\hline
$N_H~[10^{22}~{\rm cm}^{-2}]$ & $3.79_{-0.14}^{+0.07}$ &
$3.69_{-0.13}^{+0.13}$ & $3.72_{-0.09}^{+0.09}$  &
$3.97_{-0.15}^{+0.73}$ & $3.78_{-0.05}^{+0.05}$ &
$4.22_{-0.10}^{+0.04}$ \\ 
$kT_e$~[keV] & $0.76_{-0.01}^{+-0.03}$ & $0.76_{-0.02}^{+0.02}$ &
$0.76_{-0.03}^{+0.03}$ & $0.54^{+0.03}$ & $0.76_{-0.01}^{+0.03}$ &
$1.37_{-0.03}^{+0.17}$ \\  
$kT_{\rm shock}$~[keV] & $\cdots$ &  $\cdots$ & $\cdots$ &
$0.58^{+0.16}$ & $\cdots$ & $0.51_{-0.07}^{+1.15}$ \\ 
$\langle kT \rangle$~[keV] & $\cdots$ & $\cdots$ &
$1.1_{-0.12}^{+0.26}$ & $\cdots$ & $\cdots$ & $\cdots$ \\
$[$Mg$]$ & $1.69_{-0.29}^{+0.23}$ & $1.71_{-0.28}^{+0.31}$ & $\equiv
1$ & $\equiv 1.7$ & $1.68_{-0.21}^{+0.23}$ & $\equiv 1$ \\ 
$[$S$]$ & $0.58_{-0.07}^{+0.07}$ & $0.62_{-0.07}^{+0.07}$ &
$0.58_{-0.07}^{+0.07}$ & $0.68_{-0.37}^{+0.28}$ & $0.56_{-0.06}^{+0.08}$ &
$0.69_{-0.08}^{+0.15}$ \\  
$\tau$~[cm$^{-3}$~s] & $>2\times10^{12}$ & $\cdots$ &
$2.7_{-0.3}^{+0.4} \times 10^{11}$ & $7.7_{-1.8}\times10^{11}$ &
$5\times10^{13}$ & $6.3_{-0.6}^{+23} \times 10^{10}$  \\ 
$K$ & $0.038_{-0.003}^{+0.002}$ & $0.032_{-0.002}^{+0.003}$ &
$0.036_{-0.003}^{+0.003}$ & $0.050_{-0.012}^{+0.009}$ &
$0.038_{-0.004}^{+0.002}$ & $0.071_{-0.005}^{+0.001}$ \\  
\multicolumn{7}{c}{\it Foreground Source} \\
$kT$~[keV] & $0.65_{-0.15}^{+0.13}$ & $0.66_{-0.15}^{+0.12}$ &
$0.66_{-0.14}^{+0.12}$ & $0.67_{-0.66}^{+14}$ & $0.65_{-0.16}^{+0.14}$
& $0.83_{-0.12}$ \\   
K [$\times10^{-5}$] & $1.11_{-0.29}^{+0.41}$ & $1.12_{-0.29}^{+0.29}$
& $1.13_{-0.29}^{+0.29}$ & $1.13_{-0.50}^{+1.43}$ &
$1.06_{-0.22}^{+0.40}$ & $1.81_{-0.62}^{+0.60}$ \\   
\hline
$\chi^2$ & 1290.86 & 1242.07 & 1273.04 & 1813.52 & 1291.76 & 1848.89  \\ 
d.o.f & 1628 & 1629 & 1628 & 1628 & 1628 & 1628  \\
\hline
\hline
\end{tabular}
\end{center}
\vspace*{-0.5cm}
\caption{Results from jointly fitting the X-ray spectrum of Kes 17
  measured by all three detectors of {\it Suzaku}.  The first row
  indicates the model used to fit the emission from SNR Kes 17, in
  each case it was multiplied by the {\tt phabs} model to account for
  photoelectric absorption.  For all fits the foreground emission
  mentioned in \S\ref{xray} is modeled using a $N_H\equiv0$
  Raymond-Smith plasma with solar abundances.  $\tau$ indicates the
  ionization timescale (defined in \S\ref{environment}), and $K$ the
  normalization (defined in \S\ref{environment};
  \citealt{arnaud96}).  For the {\tt vgnei} model, $\langle kT
  \rangle$ is ionization timescale averaged plasma temperature.  For
  the {\tt vpshock} and {\tt vnpshock} models, $\tau$ is the highest
  ionization timescale in the plasma -- the lowest value was fixed at
  $\tau_l\equiv 0$.  The abundance of Mg in the {\tt vgnei} model was
  fixed to solar since this was the preferred value when it was
  allowed to vary.  For the {\tt vnpshock} and {\tt vsedov} models,
  the fits preferred unphysical Mg abundances, so it was fixed to the
  quoted values.  Where given, errors denote the 90\% confidence
  region, otherwise the model was not able to constrain the value of
  this parameter.  Lastly, ``d.o.f'' stands for degrees of freedom.}  
\label{tab:xrayspec}
\end{table*}

Due to the arcminute angular resolution of {\it Suzaku}, our spectra
are contaminated by unrelated objects in the source region.  Our
analysis of the previous {\it XMM} observation indicates three bright
soft X-ray sources, most likely foreground stars, located within the
{\it Suzaku} source region -- one $\sim30\times$ brighter than the
other two combined.  Our spectral analysis of their combined emission
finds it is well reproduced by a Raymond--Smith plasma with
$N_H\equiv0$ and solar \citep{anders89} abundances.  Since emission
from these sources are in the {\it Suzaku} spectra, we included such a
component in all the spectral fits described below, with the
temperature and normalization as free parameters.  For each fit, the
properties of this component are consistent with that measured for the
brightest star in the {\it XMM} data -- confirming the foreground
origin of this emission.

As shown in Figure \ref{fig:model1}, there are several prominent lines
in the X-ray spectrum of Kes 17, most notably Mg {\sc xi} and possibly
Mg {\sc xii} at $\sim1.5$~keV, Si {\sc xiii} and Si {\sc xiv} at
$\sim2$~keV, and S {\sc xv} at $\sim2.5$~keV, indicative of thermal
emission.  For most thermal models, assuming the emitting plasma has
solar abundances results in fits which under-predict the flux of these
Mg lines and over-predict the flux of these S lines.  To investigate
if these discrepancies result from uncertain calibration of the {\sc
  xis1} detector around the Si~K line, we re-fit the data excluding
these channels assuming solar abundances.  However, the same feature
was observed in the residuals of {\sc xis0} and {\sc xis3} data.
Therefore, we allowed the abundance of both Mg and S to vary in our
fits.

As shown in Table \ref{tab:xrayspec}, modeling the thermal X-ray
emission with a non-equilibrium ionization model (e.g., the {\tt vnei}
model in {\tt XSpec} and {\tt Sherpa}) results in an excellent fit
(reduced $\chi^2 \approx 0.8$) for an electron temperature of $kT_{\rm
  e} \sim 0.7$~keV, sub-solar abundance of S but a super-solar
abundance of Mg, and a very large ionization timescale $\tau$ ($\tau
\ga 2\times10^{12}$~cm$^{-3}$~s with 90\% confidence), close to the
equilibrium ionization condition \citep{smith10}.  A large $\tau$ was
also reported by \citet{gok12} in their independent analysis of this
{\it Suzaku} observation.  Not surprisingly, modeling the observed
spectrum with the \citet{raymond77} model for a diffuse hot plasma,
which assumes ionization equilibrium, also provides a very good fit to
the data for a similar electron temperature and abundances (Table
\ref{tab:xrayspec}).  However, these models assume the plasma has a
constant, uniform temperature and a single ionization state (i.e., all
the plasma was shocked at the same time to the same temperature),
unlikely to be true for an SNR.  As a result, we also fit the observed
X-ray spectrum of Kes 17 with more physically motivated models which
allow for a range of temperatures but a single ionization state ({\tt
  vgnei} and {\tt vsedov}) or a single temperature but a range of
ionization timescales ({\tt vpshock} and {\tt vnpshock}).  Of these
more physically motivated models, only {\tt vgnei} was able to
reproduce the observed spectrum (Table \ref{tab:xrayspec}).  This
model does not require a super-solar abundance of Mg, and prefers an
ionization timescale $\tau \sim 3\times10^{11}$~cm$^{-3}$~s,
significantly lower than that required by the other models (Table
\ref{tab:xrayspec}).

The success of purely thermal models in reproducing the observed X-ray
spectrum stands in contrast to the past analyses of \citet{gok12} and
\citet{combi10} which require substantial non-thermal emission.  In
their analysis of the same {\it Suzaku} data, \citet{gok12} reproduce
the electron temperature and Mg and S abundances given in Table
\ref{tab:xrayspec} but they require a power-law component with photon
index $\Gamma \sim 1.4$.  (They use a thermal model which assumes
ionization equilibrium, consistent with the Raymond-Smith model
described above.)  The power-law component is motivated by an excess
of emission below 1.5~keV resulting from fitting the observed spectrum
with a single thermal model (Figure 3 in \citealt{gok12}).  In our
spectral fits, this energy range is dominated by the foreground
component described above.  The similarity between the properties of
our foreground component and the X-ray spectral properties of these
stars measured by our analysis of the {\it XMM} observation
establishes a strong case for a purely thermal description of the {\it
  Suzaku} X-ray emission from Kes 17.

This does not explain the non-thermal X-ray emission claimed by
\citet{combi10} in their analysis of the {\it XMM} data.  These
authors divided Kes 17 into three spatial regions (none of which
included the foreground stars mentioned above) and required a
significant power-law component with $\Gamma \sim 1-3$ to reproduce
the observed emission $>4$~keV (Figure 2 in \citealt{combi10}) in each
region.  Our analysis of the {\it XMM} data following their procedure
confirms this result.  Even if one does not divide the observed X-ray
emission of Kes 17 into three spatial regions, a non-thermal component
is still needed to explain the X-ray spectrum measured by {\it XMM} --
fitting the composite X-ray spectrum of Kes 17 as measured with {\it
  XMM} with a single absorbed Raymond-Smith plasma with non-solar Mg
and S abundances\footnote{The best fit parameters are a
  $N_H=3.3^{+0.1}_{-0.1}\times10^{22}$~cm$^{-2}$ and
  $kT=0.80_{-0.03}^{+0.03}$~keV (errors denote the 90\% confidence
  interval), and this fit had a $\chi^2=1674.91$ in 1607 degrees of
  freedom.} systematically underpredicts the flux $>4$~keV.  Adding a
power-law component to this model improves the fit, reducing the
$\chi^2$ to 1665.22 with 1605 degrees of freedom.  This power-law
component has a normalization $K_{\rm PL} = 2.0_{-1.9}^{+6.3} \times
10^{-2}$~photons~cm$^{-2}$~s$^{-1}$~keV$^{-1}$ at 1~keV (errors denote
the 90\% confidence interval) and a photon index
$\Gamma=7.5_{-3.6}^{+2.3}$.  These are marginally consistent with the
analysis of \citet{combi10}, who report a total $K_{\rm PL} \sim
10^{-3}$~photons~cm$^{-2}$~s$^{-1}$~keV$^{-1}$ at 1~keV a photon index
$\Gamma=3.1\pm0.3$ in the Northern region which they claim dominates
the non-thermal X-rays emission.  This photon index is significantly
softer than the non-thermal X-ray emission detected from other SNRs
(e.g., \citealt{reynolds08}).  However, according to this f-test, the
improvement in $\chi^2$ by adding a power-law component has a
$\sim1\%$ chance of resulting from chance, and is therefore has
$\lesssim 3\sigma$ significance.

To determine if the non-thermal X-ray emission reported by
\citet{combi10} is consistent with our analysis of the significantly
deeper {\it Suzaku} data, we used {\tt XSpec} to simulate the expected
spectrum of the foreground component and an absorbed Raymond-Smith
plus power-law component with a given photon index $\Gamma$ and
normalization $K_{\rm PL}$, and then fit it using the Raymond-Smith
model + foreground component described above.  The resultant
upper-limit on $K_{\rm PL}$ is the highest value of $K_{\rm PL}$ for
which our purely thermal model was able to fit the simulated spectrum
with a reduced $\chi^2 < 2$.  Since \citet{combi10} claim the photon
index $\Gamma$ of the power-law emission varies between $\Gamma \sim 2
- 3$ in different regions, we determined the upper-limit on $K_{\rm
  PL}$ for both $\Gamma=2$ and $\Gamma=3$.  For $\Gamma=2$, we require
$K_{\rm PL}<1.5\times10^{-4}$, while for $\Gamma=3$, $K_{\rm
  PL}<5\times10^{-4}$.  Both upper-limits are inconsistent with the
results of \citet{combi10}, whose fits to the South, Center, and North
regions of Kes 17 required a combined $K_{\rm PL} \sim 10^{-3}$.  Due
to the fairly low statistical significance of the non-thermal
component in the composite SNR spectrum extracted from the {\it XMM}
data, and its $\sim2-5\times$ higher flux than allowed in the spectrum
extracted from the {\it Suzaku} observation which detected
$\sim3\times$ more photons from Kes 17 than {\it XMM}, we conclude
there is no significant non-thermal X-ray emission detected from Kes
17.

While {\it Suzaku} does not have the angular resolution to directly
detect spatial variations in the X-ray emission of Kes 17, it is
possible to use this data set to test such claims \citep{combi10}.  If
correct, modeling the observed {\it Suzaku} spectrum with the three
absorbed {\tt pshock} + power-law models used by \citet{combi10} plus
the foreground component discussed above should result in a better fit
than the single thermal models used above.  This was not the case.
However, despite having fewer degrees of freedom, the resulting fit
had a $\chi^2$ worse than the spectral fits reported in Table
\ref{tab:xrayspec}, even when we fixed the values of $N_H$, $kT$,
Abundance, $\tau$, and $\Gamma$ of each {\tt pshock} + power-law
component to those reported by \citet{combi10} (we allowed the
normalizations to vary to account for the different size extraction
regions).  Therefore, we conclude that the spatial variations in the
X-ray spectrum of Kes 17 reported by \citet{combi10} are inconsistent
with the global spectrum of this remnant, and are likely the result of
a combination of the different spatial and spectral resolutions of
{\it XMM} and {\it Suzaku} and systematic and statistical uncertainty
in the {\it XMM} background.

\begin{figure}[tb]
\begin{center}
\includegraphics[width=0.95\columnwidth]{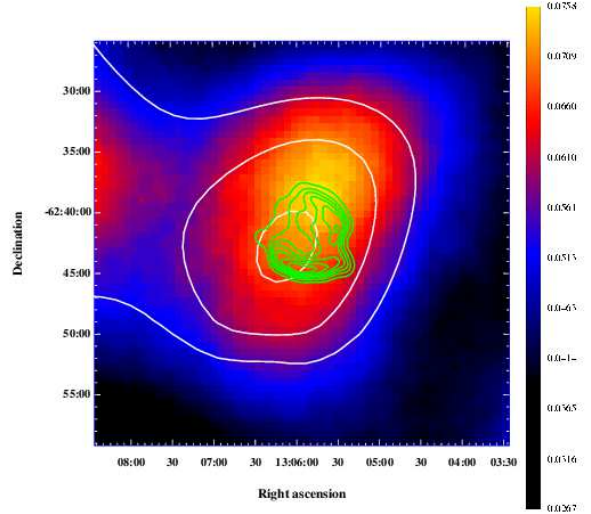}
\end{center}
\vspace*{-0.5cm}
\caption{Smoothed {\it Fermi} LAT TS map of {\it front}
converted events in the range 2 to 200 GeV of the $0\fdg6 \times
0\fdg6$ region, centered on SNR \snr. The pixel binning is $0\fdg01$,
and the maps are smoothed with Gaussians of width $0\fdg2$. Green
contours represent the radio emission (0.843 GHz) from MOST
observations. Test statistics are shown as white contours
(81-100-121).}
\label{fig:LATimg}
\end{figure}

\subsection{Gamma-rays}
\label{gray}
Kes 17 is also detected at GeV $\gamma$-ray energies \citep{wu11}.  To
determine its properties, we analyzed 39 months (from 2008 August
until 2012 February) of data collected by the {\it Fermi Gamma-ray
  Space Telescope} Large Area Telescope ({\it Fermi}-LAT).  We only
include events belonging to the Pass 7 V6 {\it Source} class, which
reduces the residual background rate (Abdo et al.\ in prep), in this
analysis.  We also use the updated (``Pass7 version 6'';
\citealt{rando09}, Abdo et al.\ in prep) instrument response functions
(IRFs), and reduce the contribution from terrestrial albedo
$\gamma$-rays by setting a maximum zenith angle for incoming photons
to 100$^{\circ}$ \citep{abdo09}.  We used the Fermi Science Tools
v9r23p1\footnote{The Science Tools package and related documentation
  are distributed by the Fermi Science Support Center at
  http://fermi.gsfc.nasa.gov/ssc}, and employed the maximum likelihood
fitting technique to analyze the morphological and spectral
characteristics of the $\gamma$-ray source \citep{mattox96}. We model
the diffuse background emission in {\it gtlike} with a Galactic
component resulting from interactions of cosmic rays with both the ISM and
photons, and isotropic components accounting for extragalactic and
residual backgrounds. The mapcube file {\tt gal\_2yearp7v6\_v0.fits}
is used to describe the $\gamma$-ray emission from the Milky Way, and
the isotropic component is modeled using the {\tt iso\_p7v6source.txt}
table.

\begin{figure}[tbh]
\begin{center}
\includegraphics[width=0.95\columnwidth]{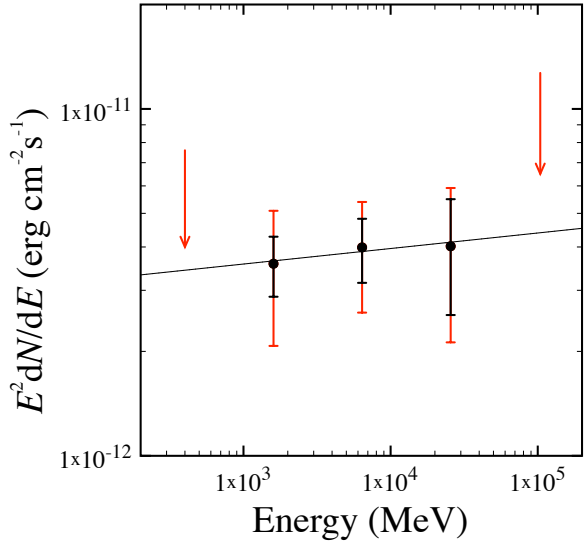}
\end{center}
\vspace*{-0.5cm}
\caption{The spectral energy distribution of the $\gamma$-ray source
  coincident with Kes 17.  The arrows represent the 95\% confidence
  intervals at these energies.  The black error bars represent
  statistical uncertainties (1$\sigma$ estimates based on
  inverse-Hessian at the optimum of the log-likelihood surface).  The
  red error bars represent systematic uncertainties, which are the sum
  in quadrature of the uncertainty related to the instrument response
  functions (IRFS), which we get from the instrument team (as cited in
  \S\ref{gray}) and the uncertainty related to variations of the
  galactic diffuse background intensity, derived by changing (and
  fixing) the normalization of the galactic background component in
  the source library for the fit to 94\% and 106\% of the best fit value
  obtained from the fit to the data at a given energy bin.
}   
\label{fig:LATspec}
\end{figure}

The spatial characteristics of the $\gamma$-ray emission in the field
of \snr\ were studied using photons between 2 and 200 GeV converted in
the {\it front} section of the LAT. For this subset of the
$\gamma$-ray data, the 68\% containment radius angle for normal
incidence photons is $\leq 0\fdg3$. We constructed test
statistic\footnote{ The test statistic is the logarithmic ratio of the
  likelihood of a point source being at a given position in a grid
  $L_{\rm ps}$, to the likelihood of the model without the additional
  source $L_{\rm null}$, $2\rm{log}(L_{\rm{ps}}/L_{\rm{null}})$.} (TS)
maps accounting for the Galactic and isotropic backgrounds using {\it
  gttsmap} and used this map to determine the statistical
significance, the position, and the possible extent of the source.  As
shown in Figure \ref{fig:LATimg}, an $\sim 11.1\sigma$ (peak
TS~$\approx124$) unresolved ($95\%$ confidence radius $=4\farcm2$)
$\gamma$-ray source with centroid $(\alpha_{\rm{2000}},
\delta_{\rm{2000}} = 13^{\rm{h}}06^{\rm{m}}05^{\rm{s}},
-62^{\circ}42'54'')$ is coincident with the radio emission.  The
residual TS map, built by modeling a point source at the best-fit
centroid of emission, shows no evidence that the source is spatially
extended since residual TS values are $<3\sigma$ within 1$^\circ$ of
the centroid.

We determined the spectral energy distribution (SED) of the
$\gamma$-ray source associated with \snr\ using data from photons with
energy between 0.2 and 204.8 GeV converted in both the {\it front} and
{\it back} sections.  We excluded photons below 200 MeV since, in this
energy range, the effective area of the instrument changes rapidly and
there are large uncertainties related to the Galactic diffuse
model. We used {\it gtlike} to model the flux in each energy bin and
estimated the best-fit parameters through the maximum likelihood
technique. To model the background in the likelihood fits we include
sources from the 24-month {\it Fermi} LAT Second Source Catalog
\citep{nolan12}\footnote{The data for the 1873 sources in the {\it
    Fermi} LAT Second Source Catalog are made available by the Fermi
  Science Support Center at {\tt
    http://fermi.gsfc.nasa.gov/ssc/data/access/lat/2yr\_catalog/}}. The
``Pass7 version 6'' IRFs we used have energy dependent systematic
uncertainties in the effective area: 10\% at 100 MeV, decreasing to
5\% at 560 MeV, and increasing to 20\% at 10 GeV (\citealt{porter09},
Abdo et al.\ in prep and references therein).  We also approximated
the effect of an uncertain underlying Galactic diffuse level by
artificially varying the normalization of the Galactic background by
$\pm6\%$ from the best-fit value at each energy bin, similar to the
analysis of \citet{castro10}.  As shown in Figure \ref{fig:LATspec},
for energies $<800$ MeV and $>51.2$ GeV only flux upper limits are
determined from the data.  The resultant SED is well-described by a
power law with spectral index of $\Gamma = 2.0 \pm 0.3$ and an
integrated photon flux above 100 MeV of $F_{>100\rm{ MeV}}\approx
1.6\times10^{-8}$ photons/cm$^{2}$/s -- similar to that measured by
\citet{wu11}.

\section{Interpretation}
\label{interpretation}
As described in \S\ref{intro}, by studying the broadband emission from
an SNR it is possible to determine the properties of both the material
inside the remnant and in the surrounding ISM.  We first analyze the
non-thermal emission observed from Kes 17 to determine the physical
origin of its $\gamma$-ray emission (\S\ref{gammaint}), and then use
those results to estimate the age of this SNR and the nature of its
environment (\S\ref{environment}).  Since the shell-like radio
morphology of Kes 17 suggests this emission originates at or near the
forward shock (\S\ref{radio}, Figure \ref{fig:20cmimg}), we assume
this SNR has a radius of $R_{\rm snr} \approx 10d_{10}~{\rm pc}$ for a
distance $d=10d_{10}~{\rm kpc}$.

\begin{figure}
\begin{center}
\includegraphics[width=0.9\columnwidth]{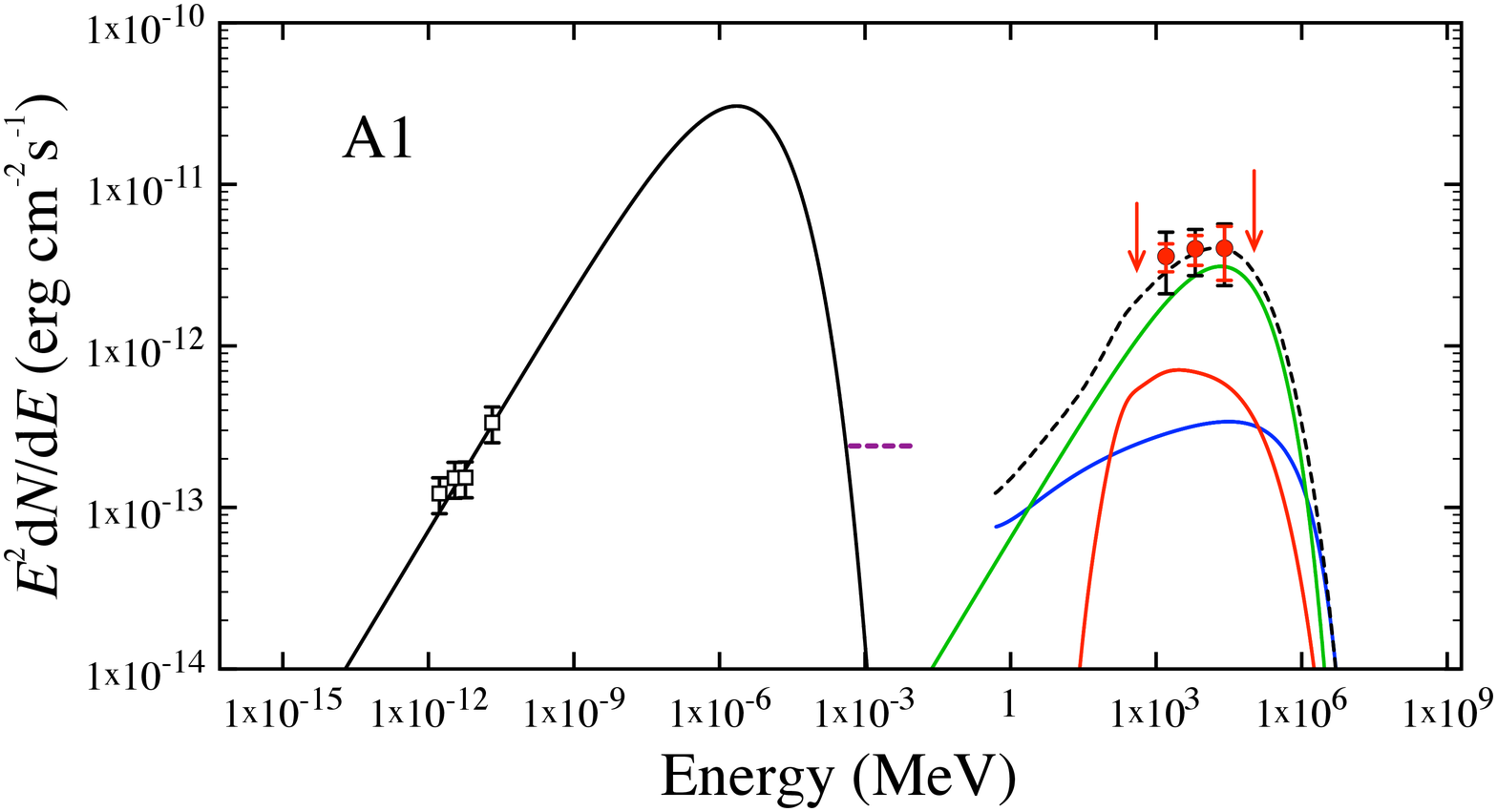}
\includegraphics[width=0.9\columnwidth]{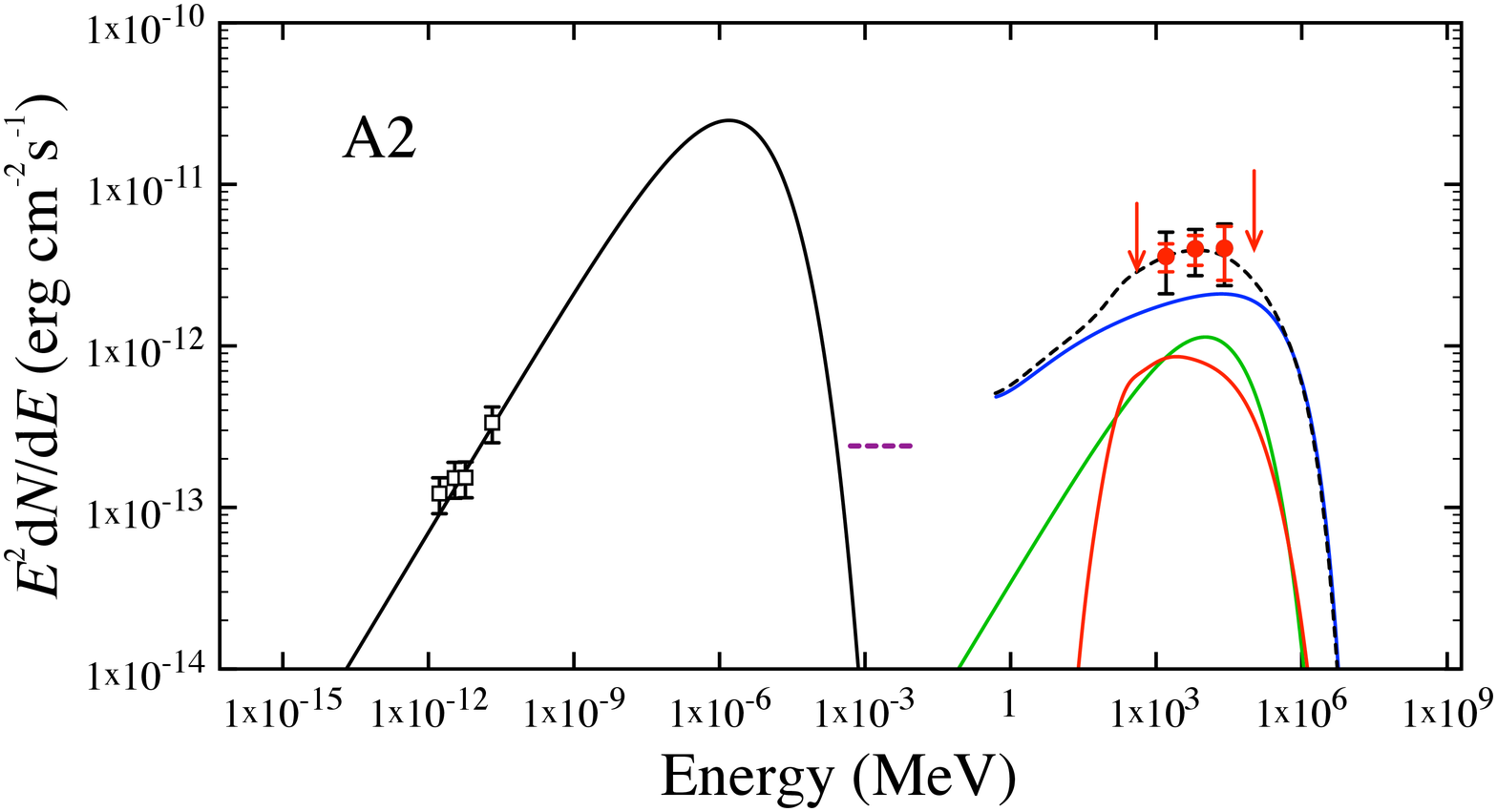}
\includegraphics[width=0.9\columnwidth]{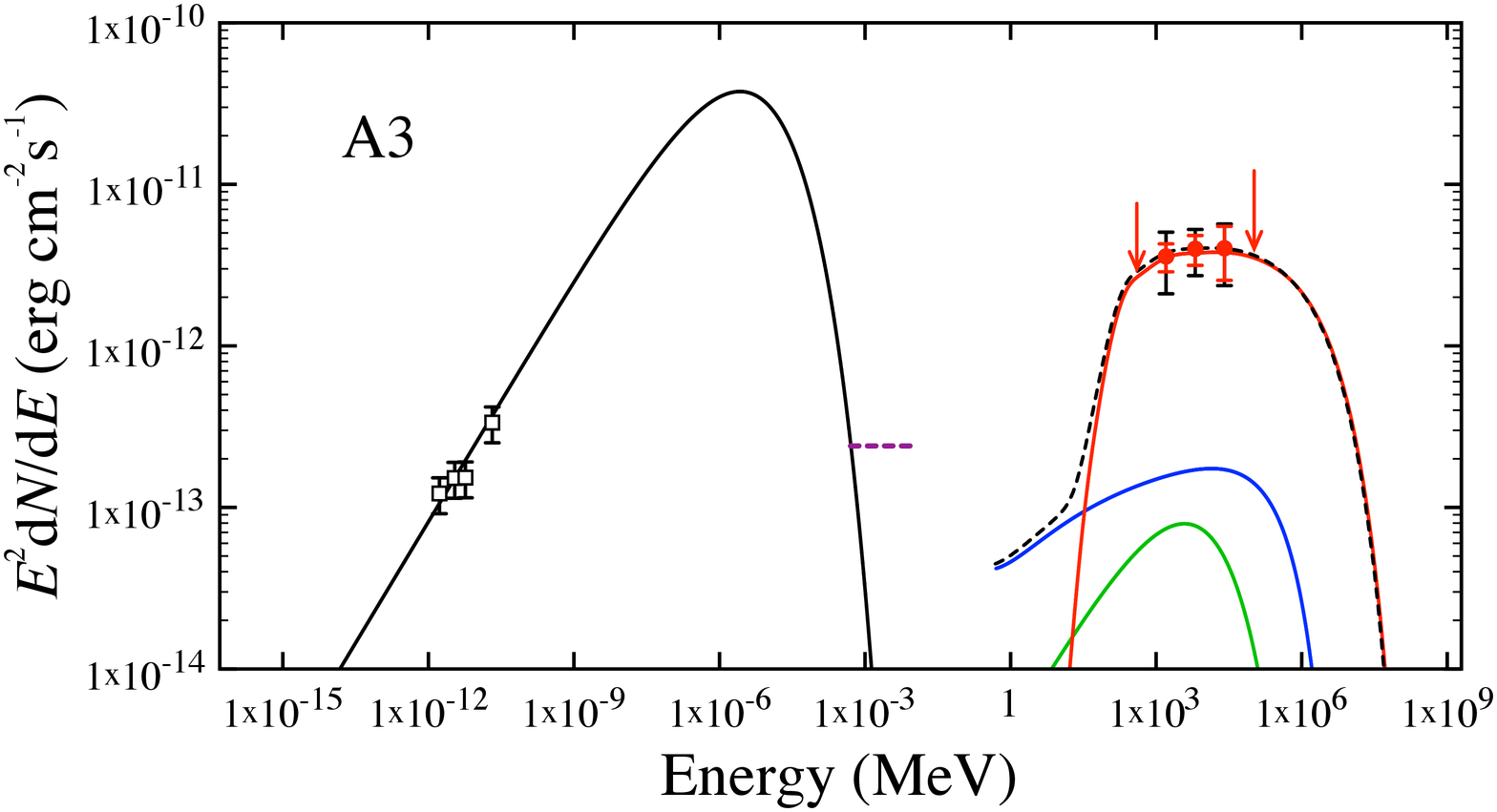}
\caption{Broadband fits to radio ({\it open squares};
  \citealt{shaver70, whiteoak96}, and {\it Fermi}-LAT ({\it red
    circles}) observations of \snr\ with the A1 ({\it top}), A2 ({\it
    middle}), and A3 ({\it bottom}) models. The modeled spectra from
  synchrotron emission ({\it black}), inverse Compton emission ({\it
    green}), $\pi^0$-decay ({\it red}), and non-thermal bremsstrahlung
  ({\it blue}), are shown. The dashed purple line indicates the upper
  limit for non-thermal X-ray emission determined from the Suzaku
  observations, using a power-law model with index $\Gamma=2$.}
\label{fig:allspec}
\end{center}
\end{figure}

\subsection{Origin of the $\gamma$-Ray Emission from Kes 17}
\label{gammaint}
Determining the properties of electrons and protons accelerated by
this SNR requires modeling the broadband spectral characteristics of
its non-thermal emission.  We assume the non-thermal radio emission is
electron synchrotron radiation, while the GeV $\gamma$-ray emission is
a combination of inverse-Compton (IC) scattering of ambient photons by
energetic electrons, non-thermal (NT) bremsstrahlung, and the decay of
$\pi^0$s produced in collisions between high energy hadrons (primarily
protons) and lower energy protons.

Fitting the observed flux densities $S_\nu$ at different radio
frequencies (Table \ref{tab:radiofluxes}) to a power law ($S_\nu
\propto \nu^\alpha$) suggests a radio spectral index $\alpha \approx
-0.6$.  We further constrain our fits using the upper-limits on
non-thermal X-ray emission derived in \S\ref{xray}.  We consider three
scenarios for the origin of the observed GeV $\gamma$-rays, each with
a different dominant emission mechanism: IC emission, NT
bremsstrahlung, or $\pi^0$-decay. We assume the spectral distribution
$dN_{p,e}/dE$ of particles accelerated in Kes 17 is:
\begin{eqnarray}
\frac{dN_{p,e}}{dE} & = & a_{p,e} E^{-\Gamma_{p,e}}
\exp\left[-\frac{E}{E_{0\,p,e}}\right],
\label{eq:uno}
\end{eqnarray}
where $E_{0\,p,e}$ is the proton/electron energy cutoff (e.g.,
\citealt{reynolds08, castro10}). The electron to proton ratio at
relativistic energies is given by the normalization coefficients of
the distributions of these particles, $K_{ep}\equiv a_e/a_p$. These
coefficients are obtained by setting the total integrated energy in
accelerated particles inside the SNR shell equal to $E_{\rm
  cr}=\eta_{\rm cr} E_{\rm sn}$, where $\eta_{\rm cr}$ is the average
efficiency of the shock in depositing energy into cosmic ray protons
and $E_{\rm sn}$ is the initial kinetic energy of the supernova
ejecta. In all models we assume $E_{\rm sn}=10^{51}$ erg, $\Gamma_ p=
\Gamma_e = 2.0$ (both the index predicted by basic Fermi acceleration
and the value derived from fitting the observed $\gamma$-ray spectrum
with a power law model; \S\ref{gray}), and the number density of
electrons $n_e=1.23 \bar{n}$ (which corresponds to material with solar
abundances), where $\bar{n}$ is the volume-averaged number density of
protons in the ISM surrounding Kes 17.

\begin{table*}[tb]
\begin{center}
\begin{tabular}{lccccccccccc}
\hline
\hline
& $K_{ep}$ & $\bar{n}$ & $B_2$ & $\eta_{\rm cr}$ & $E_{\rm 0e}$ &
$E_{\rm 0p}$ & $F_{\rm IC}$ & $F_{\rm brem}$ & $F_{\pi}$ \\  
Model & & [cm$^{-3}$] & [$\mu$G] & & [TeV] & [TeV] &
\multicolumn{3}{c}{[$10^{-9}$ ph cm$^{-2}$ s$^{-1}$]}\\  
\hline
IC Dominated (A1) & 0.1 & 1.0 & 10 & 0.6 & 2.2 & 2.2 & 6.2 & 1.5
& 2.6 \\
NT Bremss. Dominated (A2) & 0.5 & 12 & 15 & 0.08 & 1.5 & 1.5 & 2.9 &
8.9 & 3.2 \\
$\pi^0$ Decay Dominated (A3) & 0.01 & 9.0 & 70 & 0.4 & 0.9 & 30 &
0.3 & 0.8 & 13 \\
\hline
\hline
\end{tabular}
\end{center}
\vspace*{-0.5cm}
\caption{Results from fitting to the broadband spectrum of Kes 17 when
assuming different dominant GeV $\gamma$-ray emission mechanisms.
$K_{ep}$ is electron to proton ratio at relativistic energies, $\bar{n}$
is the average density of the surrounding ISM, $\eta_{\rm cr}$ is the
efficiency of cosmic ray acceleration, $B_2$ is the magnetic
field immediately behind the shock, $E_{\rm 0e}$ is the cut-off energy
of accelerated electron, and $F_{\rm IC}$, $F_{\rm brem}$, $F_{\pi}$
are respectively the flux of inverse-Compton, non-thermal bremsstrahlung, and
$\pi^0$ decay emission $>100$~MeV.} 
\label{tab:dens}
\end{table*}

We model emission from $\pi^0$-decay using the work of
\citet{kamae06}, which includes a scaling factor of 1.85 for Helium
and heavier nuclei \citep{mori09} as described by
\citet{castro10}. The synchrotron and IC emission components follow
the models presented by \citet{baring99} and references therein, and
the NT bremsstrahlung emission is modeled using the
prescription presented by \citet{bykov00}.  We assume the dominant
photon field for IC scattering is the Cosmic Microwave Background
(CMB; $kT_{\rm{CMB}} = 2.725$~K).  We also assume Kes 17 is in the
Sedov phase of its evolution, in which the shocked material is
compressed by a factor of 4.  If the swept-up material in Kes 17 is
radiatively cooling, then it will be compressed by a factor $\gg4$.
This with significantly increase the density of ambient cosmic ray
protons swept-up by the expanding SNR, enhancing their $\gamma$-ray
emission \citep{chevalier99}, as well as the density of swept-up
ambient cosmic ray electrons, possibly enhancing their $\gamma$-ray
emission as well.  However, the ages estimated in \S\ref{environment}
suggested this is not the case.

We built scenarios where each possible $\gamma$-ray emission mechanism
(IC, NT bremsstrahlung and $\pi^0$-decay emission) dominates by
adjusting the values of $K_{ep}$, $\bar{n}$, $E_{0e}$, and post-shock
magnetic field strength $B_2$, and then fit the observed broadband
spectrum.  As shown in Figure \ref{fig:allspec}, all three dominant
GeV $\gamma$-ray emission mechanism can reproduce the broadband
spectrum of this SNR given the representative model parameters are
listed in Table \ref{tab:dens}.  However, our model requires
$K_{ep}\ga0.1$ to reproduce the observed radio flux density if IC
radiation (A1) or NT bremsstrahlung (A2) dominates the $\gamma$-ray
emission -- inconsistent with the local cosmic-ray measured value of
$K_{ep} \sim 0.01$ \citep{yoshida08}.  Therefore, we conclude that
$\pi^0$ decay is primarily responsible for GeV $\gamma$-ray emission
detected from Kes 17.  Note that we assume $K_{ep}=0.01$ for the
$\pi^0$ decay scenario; similar results are obtained for lower values
of $\eta_{\rm cr}$ and $K_{ep}$ and higher values of $\bar{n}$ since
$\eta_{\rm cr} \propto \bar{n}^{-1}$ in this $\pi^0$-decay emission
model \citep{drury94}.  As a result, our modeling requires that
$\bar{n} \ga 9$~cm$^{-3}$ and $\eta_{\rm cr} \la 0.4$ for $\pi^0$
decay to dominate the $\gamma$-ray emission from Kes 17.

$\pi^0$ decay is the dominant $\gamma$-ray emission mechanism even if
the claims of non-thermal X-ray emission by \citet{combi10} or
\citet{gok12} are correct.  Modeling the observed broadband spectra
for the non-thermal X-ray fluxes reported in these papers again
requires values of $K_{ep}$ considerably higher than measured locally
if IC and/or NT bremsstrahlung emission dominate at GeV energies.
These fluxes also require a higher value of $E_{0e}$, which
subsequently changes $\bar{n}$, $B_2$, and $\eta_{\rm cr}$ but, due to
degeneracies between these parameters, the net effect is uncertain.

Neglecting background photon fields other than the CMB (energy density
$u_{\rm CMB}=0.260$~eV~cm$^{-3}$) can lead us to overestimate $K_{\rm
  ep}$ when considering IC radiation dominated $\gamma$-ray emission .
The location of Kes 17 in the Galactic plane suggests the presence of
more energetic photon fields, whose inclusion would decrease the
required value of $K_{\rm ep}$.  While IR emission is detected from
Kes 17 itself (\S\ref{ir}; \citealt{lee11}), its energy density
\footnote{We calculate $u\sim8\times10^{-4}$~eV~cm$^{-3}$ for the $T
  \approx 79$~K, $L\approx 1500~L_\odot$ modified blackbody and
  $u\sim6.5\times10^{-3}$~eV~cm$^{-3}$ for $T\approx 27$~K, $L\approx
  12500~L_\odot$ modified blackbody.} is far too low to significantly
modify the value of $K_{\rm ep}$.

Other possible photon fields are ambient starlight ($T\sim 5000$~K)
and emission from warm dust ($T \sim 25$~K).  To determine if they
could allow IC to dominate the observed $\gamma$-rays, we estimate the
photon energy density required for IC scattering of relativistic
electrons in the SNR off these photons to be primarily responsible for
the bulk of the $\gamma$-ray emission if $K_{\rm ep}=0.01$.  Since the
required photon energy density decreases for larger energy in
relativistic electrons $E_{\rm elec}$, we use the maximum electron
energy allowed by the data.  A cosmic ray acceleration efficiency
$\eta=0.4$ (as suggested by the $\pi^0$ decay model) suggests $E_{\rm
  cr} = 0.4\times10^{51}$~ergs.  For $K_{\rm ep}=0.01$ and the
electron cut-off energies given in Table \ref{tab:dens}, the total
energy in relativistic electrons is $E_{\rm elec} =
3.4\times10^{48}$~ergs.  By choosing such a high cosmic ray
acceleration efficiency, we likely overestimate the true energy in
relativistic electrons, and therefore our analysis {\it
  underestimates} the required energy density.

If the background photons are dominated by emission from warm dust, it
must have an energy density $u_{\rm dust} \ga 40u_{\rm CMB}$.  If the
background photons are dominated by starlight, it must have an energy
density $u_{\rm starlight} \ga 500u_{\rm CMB}$.  Since a combination
of the two is likely, we also fit for the required energy density of
starlight assuming $u_{\rm dust} = 20u_{\rm CMB}$.  In this case, the
required $u_{\rm starlight} \ga 200u_{\rm CMB}$.  In each scenario,
the required energy density of both dust emission and starlight is
significantly higher the values estimated from modeling the observed
diffuse $\gamma$-ray emission of the Milky Way \citep{strong00}.  As a
result, photons from starlight and warm dust incident on Kes 17 are
unlikely to have the high energy densities needed for IC emission to
be the dominant $\gamma$-ray emission mechanism.  Therefore, we
conclude $\pi^0$ decay is likely responsible for the bulk of the
$\gamma$-ray emission observed from Kes 17.

If correct, then Kes 17 is one of few ($\la10$) SNRs with direct
observational evidence for acceleration protons to high energies.  In
Table \ref{tab:snrcomp}, we compare the physical properties of Kes 17
with those other such remnants.  For many of these SNRs, the broadband
spectral modeling also assume that $K_{ep} \sim 0.01$.  Due to the
lack of spectral information at TeV energies, we can only constrain
$E_{0p} > 500$~GeV for Kes 17 (Table \ref{tab:dens}) -- higher than
the observed value for SNR IC 443, but consistent with the measured
value of other SNRs (Table \ref{tab:snrcomp}).  The strength of the
post-shock magnetic field, $B_{\rm snr}$, is also within the range
spanned by cosmic-ray producing SNRs -- though closer to the value
inferred for the older ($t_{\rm age} \ga 4000$~years) SNRs (e.g., IC
443, W51C, W28) than the younger ($t_{\rm age} < 1000$~years) SNRs
(e.g., Cas A and Tycho's SNR) among this group.  The allowed average
density $\bar{n}$ is within the range spanned by this group, though
the average cosmic-ray acceleration efficiency $\eta_{\rm cr}$
required for the lowest allowed value of $\bar{n}$ is quite high:
$\ga5\times$ higher than that of the younger SNRs and $\sim2\times$
higher than any of the older cosmic-ray producing SNRs.  Only by
determining the physical properties of the environment surrounding Kes
17, as we do in \S\ref{environment}, can we determine if this SNR is
an especially efficient producer of cosmic rays.

\subsection{Environment of Kes 17}
\label{environment}
As discussed in \S\ref{gammaint}, a primarily hadronic origin for the
GeV $\gamma$-rays detected from Kes 17 requires this SNR is expanding
into an ISM with a volume-averaged density of $\bar{n} \ga 9(\eta_{\rm
  cr}/0.4)^{-1}$~cm$^{-3}$ (\S\ref{gammaint}, Table \ref{tab:dens}).
In this section, we wish to determine if this environment is
consistent with the radius (\S\ref{radio}), dust mass (\S\ref{ir}),
and electron temperature (\S\ref{xray}), observed from this remnant.
This analysis also allows us to infer the basic properties, e.g., its
age $t_{\rm age}$ and current expansion velocity $v_{\rm snr}$, needed
to understand the underlying particle acceleration mechanism (e.g.,
\citealt{reynolds99, reynolds08}).

The $\ga 3-11$~M$_\odot$ of dust in the SNR shell inferred from IR
observations (\S\ref{ir}; \citealt{lee11}) is likely dominated by
pre-existing dust swept up by the expanding ejecta or dust formed
inside the remnant.  If Kes 17 is expanding into a medium with a
volume average density $\bar{n} = 10\bar{n}_{10}$~cm$^{-3}$ (where
$\bar{n}_{10} \ga 1$, \S\ref{gammaint}), then the mass of material
swept-up by the expanding supernova ejecta $M_{\rm sw}$ is:
\begin{eqnarray}
\label{msweqn}
M_{\rm sw} & = & \frac{4}{3}\pi R_{\rm snr}^3 \bar{n} m_{\rm p} \\
 & \approx & 1300\bar{n}_{10}d_{10}^3~M_\odot.
\end{eqnarray}
If this medium has a typical dust-to-gas mass ratio of $\sim0.1\% -
0.5\%$ (e.g., \citealt{pei92}), then Kes 17 has swept-up $\sim (1 -
5)\bar{n}_{10}~M_{\odot}$ of interstellar dust, comparable to the mass
estimated from observations.  If there is considerably more mass at
lower temperatures \citep{lee11}, this does not require the additional
dust was formed inside the SNR but is likely indicative of
$\bar{n}_{10} > 1$ and/or a higher dust-to-gas mass ratio, possible if
Kes 17 is expanding into a molecular cloud.  In fact, the detection of
OH (1720 MHz) maser (\S\ref{radio}) and molecular shock (\S\ref{ir})
emission indicates this SNR is expanding inside a molecular cloud
\citep{yusef-zadeh03}.  This implies the clumps observed in the IR are
likely the result of dense material inside the cloud swept-up and
shocked by the expanding ejecta (\S\ref{ir}).

While molecular clouds have a very complicated density structure
(e.g., \citealt{williams95}), one can approximate this environment as
a collection of clumps with average density $\bar{n}_{\rm clump}$ and
volume filling factor $f_{\rm clump}$ embedded in a uniform interclump
medium with density $n_{\rm ic}$ and volume filling factor $1-f_{\rm
  clump}$ \citep{chevalier99}.  In this model, $\bar{n}$ is:
\begin{eqnarray}
\bar{n} & = & \bar{n}_{\rm clump} f_{\rm clump} + n_{\rm ic} (1-f_{\rm clump}).
\end{eqnarray}
As noted in \S\ref{gammaint}, the acceleration efficiency $\eta_{\rm
  cr} \propto \bar{n}^{-1}$, and $\eta_{\rm cr} \sim 10\%$ (the value
inferred for other remnants; Table \ref{tab:snrcomp}) requires
$\bar{n} \sim 90$~cm$^{-3}$.  These parameters ($\bar{n}$, $n_{\rm
  ic}$, $\bar{n}_{\rm clump}$, $f_{\rm clump}$) have been measured for
a few molecular clouds, which find that typically $n_{\rm ic} \la
10~{\rm cm}^{-3}$, $f_{\rm clump}\sim10\%$, $\bar{n} \sim 20~{\rm
  cm}^{-3}$, and $\bar{n}_{\rm clump} \sim 200 - 1000$~cm$^{-3}$ with
considerable variation between clouds (e.g., \citealt{blitz93,
  williams95}).  From the analysis of the thermal X-ray spectrum of
Kes 17 presented in \S\ref{conduction} \& \S\ref{evaporation}, we
estimate $n_{\rm ic} \lesssim 0.4~{\rm cm}^{-3}$.  If $\bar{n} \sim
9$~cm$^{-3}$, $\bar{n}_{\rm clump}=200$~cm$^{-3}$, and $n_{\rm
  ic}\sim0.4$~cm$^{-3}$, then $f_{\rm clump} \sim 4\%$ -- consistent
with the observed values.  However, if $\bar{n} \sim 90$~cm$^{-3}$,
then $f_{\rm clump}$ is an extremely high $\sim 45\%$ for these values
of $\bar{n}_{\rm clump}$ and $n_{\rm ic}$.  But, if $\bar{n}_{\rm
  clump} \sim 1000$~cm$^{-3}$, as measured around H{\sc ii} region NGC
2244 which is inside a molecular cloud \citep{williams95}, then
$f_{\rm clump}\sim 10\%$ for $\bar{n} \sim 90$~cm$^{-3}$ and $n_{\rm
  ic}\sim0.4$~cm$^{-3}$.  The range of clump densities ($n_{\rm clump}
\sim 100 - 1000$~cm$^{-3}$ up to $n\sim10^4 - 10^6$~cm$^{-3}$)
inferred from analysis of the IR spectrum of Kes 17 (\S\ref{ir})
suggests $\bar{n}_{\rm clump} \sim 1000$~cm$^{-3}$ is plausible.

Interpreting the radius and X-ray temperature of Kes 17 requires
understanding its dynamical evolution.  A supernova ejects material of
mass $M_{\rm ej}$ and initial kinetic energy $E_{\rm sn}$ into its
surroundings.  Initially, the ejecta expands supersonically relative
to its environment, driving a shock called the ``forward shock'' into
its surroundings.  At this shock, the swept-up ambient material is
accelerated, compressed, and heated to a pressure significantly higher
than that of the expanding ejecta.  As a result, the shocked ambient
material drives a shock wave, called the ``reverse shock'', into the
expanding ejecta which decelerates, compresses, and heats this
material.  In the standard evolutionary model for SNRs (e.g.,
\citealt{chevalier77, truelove99} and references therein), at early
times the forward shock expands with a roughly constant velocity, such
that the radius of the forward shock $R_{\rm snr} \propto t$.  Since
it expands with constant velocity, the ejecta lose little kinetic
energy during this phase.  This ``free expansion'' phase ends when the
reverse shock has passed through all of the ejecta, approximately when
the mass swept-up by the forward shock $M_{\rm sw} \approx M_{\rm
  ej}$.  During this phase, commonly referred to as the ``Sedov-Taylor
phase,'' adiabatic losses are expected to dominate the energy
evolution of the ejecta, and the SNR expands as $R_{\rm snr} \propto
t^{2/5}$ (e.g., \citealt{chevalier77, truelove99} and references
therein).  When the radiative cooling time of the shocked gas is
comparable to the age of the SNR, radiative losses dominate,
significantly changing its dynamical evolution \citep{blondin98}.

However, the evolution of Kes 17 will be significantly different due
to its expansion into a clumpy molecular cloud (e.g.,
\citealt{chevalier99}) and its efficient acceleration of cosmic rays
(e.g., \citealt{ellison07, ellison10, ferrand10, castro11}).
Observational evidence that Kes 17 has evolved differently is its
``mixed morphology'' nature \citep{combi10}, defined by the observed
combination of steep-spectrum radio shell and interior thermal X-ray
emission \citep{rho98}.  The non-Sedov density and temperature profile
suggested by its center-filled thermal X-ray morphology can modify the
growth of the SNR (e.g., \citealt{white91}).  Currently, the two
leading physical explanations for mixed-morphology SNRs is that
thermal conduction drives gas heated at the forward shock to the
center (e.g., \citealt{cui92, chevalier99}) or dense clumps are
evaporating inside the remnant (e.g., \citealt{white91}).  While
neither model accurately reproduces the observed temperature and X-ray
surface brightness profiles of all mixed-morphology SNRs (e.g.,
\citealt{slane02}), we will interpret the radius of electron
temperature of Kes 17 using these models to roughly estimate its age
and environment.

\subsubsection{Thermal Conduction}
\label{conduction}
If heat conduction is primarily responsible for the thermal X-rays
observed in the center of Kes 17, then its evolution is likely similar
to the ``standard'' sequence outlined above.  As before, we
approximate the molecular cloud environment as a collection of
discrete, small, high density clumps embedded in low, constant density
interclump gas \citep{chevalier99} .  The expanding ejecta will shock
both the interclump gas and the clumps but, for the clump densities
inferred from IR observations ($n_{\rm clump} \sim 10^2 -
10^6$~cm$^{-3}$; \S\ref{ir}), the transmitted shock is too slow to
heat this material to X-ray emitting temperatures.  Therefore, the
mass of the X-ray emitting gas in Kes 17 $M_{X}$ should not exceed the
mass of swept-up interclump material $M_{\rm ic}$  $(M_{X} \leq M_{\rm
  ic})$.

It is possible to estimate $M_{X}$ from the fits to thermal X-ray
spectrum presented in \S\ref{xray}. The mass of the X-ray emitting gas
is equal to:
\begin{eqnarray}
M_X & = & \frac{4}{3}\pi R_{\rm snr}^3  f_X n_{H,X} m_p,
\end{eqnarray}
where $m_p$ is the mass of the proton, $n_{H,X}$ is the density of the
X-ray emitting gas, and $f_X$ is the fraction of the SNR's volume
filled with the X-ray emitting plasma.  We can estimate $n_{H,X}$ from
the normalization $K$ of the thermal X-ray emission models used in
\S\ref{xray}, since \citep{arnaud96}:
\begin{eqnarray}
K & = & \frac{10^{-14}}{4\pi d^2}\int n_{e,X} n_{H,X} dV \\
 & = & 0.3 f_X \mu^{-1} n_{H,X}^2 d \theta_{\rm snr}^3  \times10^{-14}
\end{eqnarray}
where $\mu$ is the number ratio of electrons to protons in the plasma
($n_{H,X} = n_{e,X} / \mu$; $\mu=1.23$ for solar abundance), $d$ is
the distance to the source in cm, and $\theta_{\rm snr}$ is the
angular radius of Kes 17 in radians.  For the measured values of $K$
(Table \ref{tab:xrayspec}, \S\ref{xray}), $n_{H,X}$ is approximately:
\begin{eqnarray}
\label{nhxeqn}
n_{H,X} & \sim & 0.4 f_X^{-\frac{1}{2}} d_{10}^{-\frac{1}{2}}~{\rm cm}^{-3}.
\end{eqnarray}
For a standard SNR, $f_X \approx 1/12$, but a mixed morphology SNR
likely has $f_X$ greater than this value.  Since $1/12 \leq
f_X \leq 1$, we therefore estimate:
\begin{eqnarray}
n_{H,X} & \sim & (0.4 - 1.4)d_{10}^{-\frac{1}{2}}~{\rm cm}^{-3}.
\end{eqnarray}

Relating $M_X$ to $M_{\rm ic}$ requires estimating the cooling time of
the X-ray emitting gas $t_{\rm cool}$.  If Kes 17 is significantly
older than the cooling time, then $M_{X} \ll M_{\rm ic}$, while if Kes
17 is younger than the cooling time, then $M_{X} \approx M_{\rm ic}$.
The cooling time can be approximated as:
\begin{eqnarray}
t_{\rm cool} & \approx & \frac{E_{\rm thermal}}{L_{\rm thermal}},
\end{eqnarray}
where $E_{\rm thermal}$ is the thermal energy of this gas and $L_{\rm
  thermal}$ is its thermal luminosity.  The thermal energy is roughly:
\begin{eqnarray}
E_{\rm thermal} & = & \frac{4}{3}\pi R_{\rm snr}^3 n_{H,X} f_X k_B T_X ,
\end{eqnarray}
where $k_B$ is Boltzmann's constant and $T_X$ is the X-ray
temperature, while the thermal luminosity $L_{\rm thermal}$ is:
\begin{eqnarray}
L_{\rm thermal} & = & \frac{4}{3}\pi R_{\rm snr}^3 \Lambda
\end{eqnarray}
where, for solar abundances and temperature $kT_X \sim 1~{\rm keV}$
(the average electron temperature and approximate chemical composition
suggested by our modeling; \S\ref{xray}), $\Lambda$ is
\citep{raymond76}:
\begin{eqnarray}
\Lambda & \sim & 5 n_{\rm e,X} n_{\rm H,X} \times 10^{-23}~{\rm
  ergs~s^{-1}~cm^{-3}} \\
& \sim & 5 \mu n_{\rm H,X}^2 \times 10^{-23}~{\rm
  ergs~s^{-1}~cm^{-3}}.
\end{eqnarray}
This suggests that:
\begin{eqnarray}
\label{eqn:tcool}
t_{\rm cool} & \approx & \frac{k_B T_x f_X}{5\mu n_{H,x} \times
  10^{-23}} \\
 & \approx & 2f_X^{\frac{1}{2}}\times10^{6}~{\rm years} \\
 & \sim & (0.6-2)\times10^{6}~{\rm years}
\end{eqnarray}
for the range of $f_X$ argued above.  This is considerably higher than
the age of the other SNRs identified as efficient cosmic ray
accelerators (Table \ref{tab:snrcomp}).  Therefore, in this scenario
it is likely that $M_X \approx M_{\rm ic}$.

The mass of the interclump gas swept-up by the expanding ejecta is:
\begin{eqnarray}
M_{\rm ic} & = & \frac{4}{3}\pi R_{\rm snr}^3 (1-f_{\rm clump}) n_{\rm
  ic} m_p.
\end{eqnarray}
For $M_X \approx M_{\rm ic}$, we have:
\begin{eqnarray}
n_{\rm ic} & \approx & \frac{f_X}{1-f_{\rm clump}} n_{H,X} \\
 & \approx & 0.4 f_X^{\frac{1}{2}} (1-f_{\rm clump})^{-1}
d_{10}^{-\frac{1}{2}}~{\rm cm}^{-3}
\end{eqnarray}
If $f_{\rm clump}\sim10\%$ as suggested by observations (e.g.,
\citealt{blitz93}, \citealt{williams95}), the allowed range of $f_X$
favors $n_{\rm ic} \sim 0.1 - 0.4~{\rm cm}^{-3}$.

As mentioned above, in this scenario the evolution of Kes 17 should be
similar to that of the ``standard'' SNR.  Therefore, its observed
radius $R_{\rm snr}$ suggests an age (e.g., \citealt{lozinskaya92}):
\begin{eqnarray}
t_{\rm age} & = & \left(\frac{R_{\rm snr}}{1.15} \right)^{\frac{5}{2}}
\left(\frac{n_{\rm ic} \mu m_{\rm p}}{E_{\rm sn}}
\right)^{\frac{1}{2}} \\
 & \approx & 4200 d_{10}^{\frac{5}{2}} \left(\frac{n_{\rm
       ic}}{0.4}\right)^{\frac{1}{2}} E_{51}^{-\frac{1}{2}}~{\rm years},
\end{eqnarray}
where $E_{\rm sn}=10^{51}E_{51}$~ergs.  For the values of $n_{\rm ic}$
estimated above, Kes 17 is only $t_{\rm age} \sim 2000 - 4200$ years
old.  This is considerably lower than the cooling timescale $t_{\rm
  cool}$ calculated above, consistent with our assumption that $M_X
\approx M_{\rm ic}$.  While this age estimate ignores the effect of
cosmic ray acceleration, numerical studies suggest that this analysis
underestimates the age by $\sim20\%$ in the case of extremely
efficient particle acceleration ($\eta_{\rm cr} \sim 40\%$; e.g.,
\citealt{castro11}).  Therefore, in this scenario we estimate that Kes
17 is $t_{\rm age} \sim 2000 - 5200$ years old.

If correct, Kes 17 is currently expanding with a speed $v_{\rm snr}$
(e.g., \citealt{lozinskaya92, truelove99}):
\begin{eqnarray}
v_{\rm snr} & \approx & 0.43 \left(\frac{E_{\rm sn}}{n_{\rm ic}m_{\rm
    p}}\right)^{\frac{1}{2}} R_{\rm snr}^{-\frac{3}{2}} \\
 & \sim & 570 \left(\frac{n_{\rm ic}}{\rm
  cm^{-3}}\right)^{-\frac{1}{2}}~\frac{\rm km}{\rm s}. 
\end{eqnarray}
If $n_{\rm ic} \sim 0.1-0.4~{\rm cm}^{-3}$ as derived above, then
$v_{\rm snr} \sim 900 - 1800~{\rm km~s}^{-1}$.  Such a shock is
expected to heat electrons to a temperature $T_e$:
\begin{eqnarray}
\label{eqn:electemp}
kT_e & \approx & \frac{3}{16} m_e v_{\rm snr}^2 \\
 & \sim & 0.9-3~{\rm eV},
\end{eqnarray}
substantially lower than the $kT_{e} \sim 0.8~{\rm keV}$ inferred from
our modeling of the observed X-ray spectrum (\S\ref{xray}).  However,
ions are heated to a temperature $T_i$:
\begin{eqnarray}
\label{eqn:iontemp}
kT_i & \approx & \frac{3}{16} m_p v_{\rm snr}^2 \\
 & \sim & 1.5-6~{\rm keV},
\end{eqnarray}
higher than the observed electron temperature.  However, many electron
heating mechanisms operate inside a SNR.  Observations suggest that,
at a forward shock expanding with $v_{\rm snr} \sim 900 - 1800~{\rm
  km~s}^{-1}$, electrons will be heated to a temperature of $T_e \sim
(0.1-0.8)T_i$ at the forward shock for this range of shock velocities
\citep{ghavamian07}, with this process possibly enhanced by efficient
particle acceleration at the forward shock \citep{castro11}.
Additionally, inside the SNR, ions heat the electrons through Coulomb
collisions.  The high ionization timescale inferred from our modeling
of the observed thermal X-ray spectrum suggests at least rough thermal
equilibration between electrons and ions in this remnant.  Therefore,
this scenario is consistent with the observed electron temperature.

In summary, if thermal conduction is the dominant mechanism
responsible for the mixed morphology nature of Kes 17, then this
remnant is $\sim2000-5000$ years old and is expanding into a clumpy
medium with an inter clump density of $\sim0.1-0.4~{\rm cm}^{-3}$.

\subsubsection{Clump Evaporation}
\label{evaporation}
Alternatively, the mixed morphology nature of Kes 17 could result from
dense clumps swept up by the expanding ejecta evaporating inside the
remnant and then being heated to X-ray temperatures by the hot
interclump gas shocked at the forward shock.  In this case, we expect
$M_X > M_{\rm ic}$.  Repeating the analysis of \S\ref{conduction},
this requires that:
\begin{eqnarray}
n_{\rm ic} & \lesssim & 0.4 f_X^{\frac{1}{2}} (1-f_{\rm clump})^{-1}
d_{10}^{-\frac{1}{2}}~{\rm cm}^{-3}.
\end{eqnarray}
Since $f_X < 1$, in this scenario $n_{\rm ic} < 0.4~{\rm cm}^{-3}$.  

The evaporation of clumps inside the SNR can significantly impact the
dynamics of the forward shock, which is now expected to expand as:
\begin{eqnarray}
R_{\rm snr} & = & \left[\frac{25(\gamma+1)\kappa E_{\rm sn}}{16\pi
    n_{\rm ic} m_{\rm p}} \right]^{\frac{1}{5}} t^{\frac{2}{5}},
\end{eqnarray}
where $\gamma=5/3$ is the adiabatic index of the surrounding material
and $\kappa$ is the ratio of thermal to kinetic energy in the SNR
\citep{white91}.  Therefore, the age of Kes 17 is:
\begin{eqnarray}
t_{\rm age} & = & \left[\frac{25(\gamma+1)}{16\pi m_{\rm p}}
  \right]^{-\frac{1}{2}} \left(\frac{\kappa E_{\rm sn}}{n_{\rm ic}}
\right)^{-\frac{1}{2}} R_{\rm snr}^{\frac{5}{2}} \\
 & \approx & 6000~\kappa^{-\frac{1}{2}} E_{51}^{-\frac{1}{2}} n_{\rm
  ic}^{\frac{1}{2}}~{\rm years} 
\end{eqnarray}
for $d_{10}=1$, where $\kappa$ can vary between 0.01 and 1
\citep{white91}.  Setting $\kappa = 0.01$ and $n_{\rm ic} = 0.4~{\rm
  cm}^{-3}$ suggests that $t_{\rm age} < 40000$ years old.  While this
analysis ignores the effect of efficient particle acceleration on the
evolution of the forward shock \citep{white91}, the resulting
$\sim20\%$ error suggested by simulations (e.g., \citealt{castro11})
is considerably less than the uncertainty resulting from the unknown
values of $\kappa$ and $n_{\rm ic}$.

In this scenario, $v_{\rm snr}$ is given by:
\begin{eqnarray}
v_{\rm snr} & = & \frac{2}{5} \left[\frac{25(\gamma+1)\kappa E_{\rm sn}}{16\pi
    n_{\rm ic} m_{\rm p}} \right]^{\frac{1}{5}} t_{\rm
  age}^{-\frac{3}{5}}, \\
 & \approx & 120,000 \left(\frac{\kappa E_{\rm 51}}{n_{\rm ic}}
\right)^{\frac{1}{5}} \left(\frac{t_{\rm age}}{\rm 1~year}
\right)^{-\frac{3}{5}}~\frac{\rm km}{\rm s}.
\end{eqnarray}
By considering the ``maximum age'' case above ($\kappa=0.01$, $n_{\rm
  ic}=0.4$~cm$^{-3}$, and $t_{\rm age}=40000$~years), we that
currently $v_{\rm snr} \geq 100$~km~s$^{-1}$.  From Equations
\ref{eqn:electemp} \& \ref{eqn:iontemp}, this minimum velocity is too
slow to currently heat either electrons or ions to the measured
electron temperature $kT_{e}\sim0.8~{\rm keV}$ (\S\ref{xray}).  This
can be rectified by Kes 17 either being younger than the maximum age
estimated above, suggesting $\kappa > 0.01$ and/or $n_{\rm ic} <
0.4~{\rm cm}^{-3}$, or the emitting material was heated at an earlier
time when the SNR was expanding faster.  This is plausible since the
time required for the X-ray emitting gas to cool ($t_{\rm cool}
\gtrsim 6\times10^5~{\rm years}$; \S\ref{conduction}, Equation
\ref{eqn:tcool}) is longer than the maximum age of Kes 17 in this
scenario.  Therefore, it is plausible that material heated at earlier
times would still radiate today.

In summary, if clump evaporation is the dominant cause of the mixed
morphology nature of Kes 17, it is expanding into a medium with an
interclump density $n_{\rm ic} < 0.4~{\rm cm}^{-3}$ and is $<40000$
years old.

\section{Conclusions}
\label{conclusions}

In this paper, we analyze and interpret recent observations of SNR Kes
17 across the electromagnetic spectrum (\S\ref{observations}).  Our
analysis indicates this SNR has a partial radio shell with a diameter
of $\sim7\farcm5$, which translates to a physical radius of
$R\sim10d_{10}$~pc at a distance $d=10d_{10}$~kpc.  The detection of
OH 1720~MHz maser emission and the IR spectrum of Kes 17 suggest this
SNR is expanding into a molecular cloud, and our analysis of a recent
{\it Suzaku} observation of this SNR suggests the observed X-ray
emission is predominantly thermal, emitted by a plasma with density
$n_{H,X} \sim 0.4$~cm$^{-3}$, roughly solar abundances except for an
under-abundance of S, and comprised of gas with an electron
temperature $kT_{\rm e} \approx 0.8$~keV in roughly thermal
equilibrium (\S\ref{xray} \& \S\ref{environment}; Table
\ref{tab:xrayspec}).  Lastly, our analysis of {\it Fermi} observations
of this field strongly detects GeV $\gamma$-ray emission coincident
with this remnant that is almost certainly from the SNR shell
(\S\ref{gray}).

By modeling the broadband non-thermal emission of Kes 17, we
determined that the GeV $\gamma$-rays are predominantly the result of
cosmic ray protons accelerated at the SNR's forward shock colliding
with swept-up material inside the SNR, producing $\pi^0$s which decay
into $\gamma$-rays (\S\ref{gammaint}).  This explanation requires that
this SNR is expanding into medium with an average density
$\bar{n}\ga9$~cm$^{-3}$ (\S\ref{gammaint}), consistent with the
molecular cloud environment implied by the OH maser emission and
considerable dust mass inferred from its IR spectrum
(\S\ref{environment}).  The age of Kes 17 and density of the
interclump medium inside the cloud $n_{\rm ic}$ depends on whether
thermal conduction or evaporation of dense clumps is primarily
responsible for the mixed morphology nature of this remnant.  If
thermal conduction is responsible, then Kes 17 is expanding into an
environment with $n_{\rm ic} \sim 0.1-0.4~{\rm cm}^{-3}$ and is only
$\sim2000-5000$ years old (\S\ref{conduction}).  However, if the
evaporation of density clumps is primarily responsible, then Kes 17 is
expanding inside an environment with $n_{\rm ic} < 0.4~{\rm cm}^{-3}$
and can be as much as $\sim40000$ years old (\S\ref{evaporation}).  If
the cosmic ray efficiency $\eta_{\rm cr}$ of Kes 17 is similar to that
other SNRs believed to be accelerating cosmic rays (Table
\ref{tab:snrcomp}), then it is expanding in an environment with
$\bar{n}\sim90$~cm$^{-3}$ (\S\ref{environment}).  This requires the
surrounding clumps have an average density of $\bar{n}_{\rm
  clump}\sim1000$~cm$^{-3}$ for a reasonable clump mass fraction of
$f_{\rm clump} \sim 10\%$ (\S\ref{environment}).  Such an average
clump density has been observed in some molecular clouds
\citep{blitz93} as well around massive stars which have formed a
stellar wind bubble or H{\sc ii} inside a molecular cloud
\citep{williams95}.

The possible high cosmic ray acceleration efficiency inferred in Kes
17 is very interesting -- especially given its likely expansion into a
clumpy medium.  Much about the particle acceleration mechanism inside
SNRs, particularly how its efficiency depends on its surroundings, is
unknown.  Recent theoretical work suggests that expansion into a
turbulent, clumpy, strongly magnetized environment enhances cosmic ray
acceleration (e.g., \citealt{bykov00, zhang09}), and further study of
Kes 17 would test these results.  This requires better understanding
the environment of Kes 17, specifically the clumpiness of its
surroundings.  Additional observations are also needed to measure the
properties of the accelerated cosmic rays in order to test models of
the acceleration mechanism.  This can be accomplished in a variety of
ways, e.g. measuring the TeV $\gamma$-ray spectrum will allow us to
determine the maximum energy of cosmic rays accelerated in the SNR
$E_{0p}$, mapping CO emission around Kes 17 will allow us to measure
the average density $\bar{n}$ of its environment (e.g.,
\citealt{williams95}), and a deeper X-ray observation will allow us
use the observed thermal X-ray emission (e.g., \citealt{ellison10}) to
constrain the cosmic ray acceleration efficiency $\eta_{\rm cr}$,
better constrain the energetics of any cosmic ray electrons
accelerated in this SNR, and determine the origin of its mixed
morphology appearance.  In any case, future study of Kes 17 is
extremely important for understanding how energetic particles are
accelerated in SNRs.

\acknowledgements The authors thank the anonymous referee for usual
comments and suggestions.  JDG acknowledges the support of NASA grant
NNX10AR51G, and NSF Astronomy and Astrophysics Postdoctoral Fellowship
grant AST-0702957, as well as the hospitality of the Center for
Cosmology and Particle Physics at New York University where much of
this work was conducted.  POS acknowledges support from NASA Contract
NAS8-03060.  The Australia Telescope is funded by the Commonwealth of
Australia for operation as a National Facility managed by CSIRO.  This
research has made use of data obtained from the Suzaku satellite, a
collaborative mission between the space agencies of Japan (JAXA) and
the USA (NASA).

\bibliography{ms}

\begin{table*}[b]\small
\begin{center}
\begin{tabular}{cccccccccc}
\hline
\hline
SNR & Age & $\bar{n}$ & ISM & $B_{\rm snr}$ & $\eta_{\rm cr}$ & $K_{\rm
ep}$ & $E_{0,p}$ & Citations \\
\hline
Kes 17 & $\sim$2,000$-$40,000 & $>9$~cm$^{-3}$& Clumpy & 35~$\mu$G &
$<0.4$ & 0.02 & $>500$ GeV &  $\cdots$ \\  
\hline
Cas A & 330~years & 30~cm$^{-3}$ & Clumpy & $0.5-1$~mG & $0.005 - 0.02$ &
$0.004 - 0.02$ & $10-30$ TeV & \tablenotemark{a}, \tablenotemark{b},
\tablenotemark{c}, \tablenotemark{v} \\ 
Tycho & 440~years & $\sim0.3$~cm$^{-3}$ & Uniform & $200-300~\mu$G &
$0.06-0.075$ & 0.0016 & $>470$~TeV & \tablenotemark{h},
\tablenotemark{i}, \tablenotemark{j}, \tablenotemark{k}\\ 
IC 443 & 4,000~years & $\sim250$~cm$^{-3}$ & Clumpy & 10~$\mu$G &
$0.006-0.02$ & 0.01$-$0.03 & 100$-$200~GeV &  \tablenotemark{d},
\tablenotemark{e}, \tablenotemark{f}, \tablenotemark{g} \\
W44 & 20,000~years & $\sim100$~cm$^{-3}$ & Clumpy & $40-800~\mu$G &
$0.03-0.15$ & $0.01 - 0.05$ & $\cdots$ & \tablenotemark{n},
\tablenotemark{o}, \tablenotemark{p}, \tablenotemark{q},
\tablenotemark{g}, \tablenotemark{r} \\ 
W51C & 30,000~years & 10~cm$^{-3}$ & Clumpy & $<150~\mu$G & 0.16 & 0.0125 &
120~TeV & \tablenotemark{s}, \tablenotemark{t}, \tablenotemark{g}, 
\tablenotemark{u} \\ 
W28 & 40,000~years & $\ga100$~cm$^{-3}$ & Clumpy & 40 -- 160~$\mu$G & 0.01
-- 0.03 & 0.01 & $\cdots$ & \tablenotemark{l}, \tablenotemark{m},
\tablenotemark{g} \\  
\hline
\hline
\end{tabular}
\end{center}
\tablenotetext{a}{\citet{berezhko03}}
\tablenotetext{b}{\citet{berezhko04}}
\tablenotetext{c}{\citet{abdo10d}} 
\tablenotetext{d}{\citet{troja08}}
\tablenotetext{e}{\citet{abdo10}}
\tablenotetext{f}{\citet{tavani10}}
\tablenotetext{g}{\citet{tang11}}
\tablenotetext{h}{\citet{cassam07}}
\tablenotetext{i}{\citet{eriksen11}}
\tablenotetext{j}{\citet{giordano12}}
\tablenotetext{k}{\citet{morlino12}}
\tablenotetext{l}{\citet{abdo10b}}
\tablenotetext{m}{\citet{giuliani10}}
\tablenotetext{n}{\citet{reach05}}
\tablenotetext{o}{\citet{abdo10c}}
\tablenotetext{p}{\citet{uchiyama10}}
\tablenotetext{q}{\citet{giuliani11}}
\tablenotetext{r}{\citet{uchiyama12}}
\tablenotetext{s}{\citet{koo95}}
\tablenotetext{t}{\citet{koo10}}
\tablenotetext{u}{\citet{aleksic12}}
\tablenotetext{v}{\citet{kim08}}
\vspace*{-0.5cm}
\caption{\scriptsize The physical properties of SNRs other than Kes 17
  with direct observational evidence for proton acceleration.
  $\bar{n}$ is the average density of the surrounding ISM, $B_{\rm
    snr}$ is the strength of the magnetic field inside the SNR,
  $\eta_{\rm cr}$ is the ratio between $E_{\rm cosmic ray}$ and the initial
  kinetic energy of the progenitor SN (assumed in many cases to be
  $10^{51}$ ergs), $K_{ep}$ is the relative normalization between
  accelerated electrons and positrons, and $E_{0,p}$ is the cutoff
  energy in the acceleration proton spectrum.  Ranges given for
  various values reflect differences in the literature, and for some
  SNRs the reported values are assumptions used in the modeling as
  opposed to fitted values.  No value for $E_{0,p}$ is given for SNRs
  W28 and W44 since a broken power-law cosmic-ray injection spectrum,
  as opposed to a power-law with an exponential cutoff, is needed to
  reproduce their non-thermal spectra.}
\label{tab:snrcomp}
\end{table*}

\end{document}